\documentclass[12pt]{iopart}

\usepackage{amssymb}
\usepackage{graphicx}
\usepackage{color}

\newcommand{\vgamma}{1}
\newcommand{\vm}{2}

\begin{document}

%\title[]{Absolute negative mobility in driven periodic systems: exploring the parameters space}
\title[]{Anomalous transport in driven periodic systems: distribution of the absolute negative mobility effect in the parameter space}

\author{Mateusz Wi\'{s}niewski and Jakub Spiechowicz}

\address{Institute of Physics, University of Silesia, 40-007 Katowice, Poland}
\ead{jakub.spiechowicz@us.edu.pl}
%\vspace{10pt}

\begin{abstract}
Absolute negative mobility is one of the most paradoxical forms of anomalous transport behaviour. At the first glance it contradicts the superposition principle and the second law of thermodynamics, however, its fascinating nature bridges nonlinearity and nonequlibrium in which these fundamental rules are no longer valid.  We consider a paradigmatic model of the nonlinear Brownian motion in a driven periodic system which exhibits the absolute negative mobility. So far research on this anomalous transport feature has been limited mostly to the single case studies due to the fact that this model possesses the complex multidimensional parameter space. In contrast, here we harvest GPU supercomputers to analyze the distribution of negative mobility in the parameter space. We consider nearly $10^9$ parameter regimes to discuss how the emergence of negative mobility depends on the system parameters as well as provide the optimal ones for which it occurs most frequently.
\end{abstract}

%\vspace{2pc}
%\noindent{\it Keywords}: Brownian motion, negative mobility, GPU, supercomputing

%\submitto{\NJP}

\maketitle

%\ioptwocol

\section{Introduction}
Nonlinear systems exhibit rich spectrum of unusual behaviour which is absent in their linear counterparts \cite{strogatz}. It is rooted in the fact that they are exempted from the superposition principle which loosely speaking tells that the response of linear system caused by two or more forces is the sum of the reactions that would have been induced by each of them individually. %When the system is nonlinear this fundamental law does not hold true anymore and it 
This property opens a new avenue for the emergence of remarkable phenomena like chaos, in which deterministic evolution of the system is completely not predictable \cite{ott} or multistability, when several stable states coexist in the setup dynamics \cite{pisarchik2014, spiechowicz2021pre}.

Similarly, when the system is taken out of thermal equilibrium, monumental Thermodynamic Laws and various symmetries such as the detailed balance lose their validity. Solely this remark opens a new landscape of phenomena that despite many years of active research in nonequilibrium statistical physics still remains a \emph{terra incognita}. Yet, some progress in exploring this fascinating ground has been achieved in the form of understanding effects like, for instance, stochastic resonance \cite{gammaitoni1998}, anomalous diffusion \cite{metzler2014,spiechowicz2016scirep,spiechowicz2017scirep,bialas2020}, noise assisted transport \cite{hanggi2009,cubero2016} or deriving various celebrated fluctuation theorems \cite{jarzynski2011,campisi2011,talkner2020} which bridge physics in and out of equilibrium.

In this work we unite these two worlds of nonlinearity and nonequilibrium that inspired the entire research fields which, as outlined above, have been intensively explored over recent decades. In doing so we investigate the form of anomalous transport behaviour, namely the absolute negative mobility, in which the particle moves in the direction opposite to the net acting force around zero bias \cite{eichhorn2002a, eichhorn2002b, ros2005, machura2007, speer2007, kostur2008, nagel2008, eichhorn2010, spiechowicz2013jstatmech, malgaretti2014, spiechowicz2014pre, sarracino2016, slapik2018, cecconi2018, ai2018, mukamel2018, slapik2019prl, slapik2019prappl, spiechowicz2019njp, ai2020physa, ai2020, ai2022}. This phenomenon rests on the combination of both nonlinearity as well as nonequilibrium and cannot emerge without these two properties in a one-dimensional system \cite{machura2007, speer2007}.

Therefore we consider a paradigmatic model of nonequilibrium statistical physics, namely, the nonlinear Brownian motion in a driven periodic system. Since it possesses a multidimensional parameter space which has been too complex to explore systematically, so far research on the absolute negative mobility has been limited mostly to the single case studies. In contrast, in this work we exploit the state of the art computer simulations to analyze the distribution of absolute negative mobility effect in the parameter space by harvesting the power of GPU supercomputers. This innovative method \cite{spiechowicz2015cpc} allowed us to consider nearly $10^9$ parameter regimes to draw a number of important qualitative and quantitative conclusions about the emergence of absolute negative mobility. In particular we provide parameters of the model which are optimal for the occurrence of this anomalous transport behaviour.

The paper is organized as follows. In Sec. 2 we recall the formulation of the model, introduce the dimensionless quantities and discuss differences between the two most common scaling regimes. In Sec. 3 we briefly review the state of the art of the absolute negative mobility effect. The next Sec. 4 contains the description of employed research methodology. In Sec. 5 we present the results of our simulations. First we discuss the qualitative dependence of the absolute negative mobility on the model parameters. Then we elaborate on the distribution of negative mobility effect in the parameter space. Section 6 provides a summary and conclusions.

\section{Model} \label{sec:model}
In this work we consider the Langevin equation describing the dynamics of a Brownian particle dwelling in a one-dimensional spatially periodic potential \cite{slapik2019prl}. The kinetics of the particle depends on its mass $M$ and friction coefficient $\Gamma$. We assume that the potential is symmetric and has a spatial period $L$, namely
\begin{equation}
	U(x) = \Delta U\sin{\left( 2\pi \frac{x}{L} \right)}.
\end{equation}
The particle is driven by a harmonic force $A\cos{(\Omega t)}$ as well as a constant bias $F$. The system is coupled to a thermostat of temperature $T$. Thermal fluctuations are modelled by $\delta$-correlated Gaussian white noise $\xi(t)$ of vanishing mean, i.e. 
\begin{equation}
	\langle \xi(t) \rangle = 0$,\quad $\langle \xi(t)\xi(s) \rangle = \delta(t-s).
\end{equation}
Such a model can be expressed by the following Langevin equation \cite{slapik2019prl}
\begin{equation} \label{eq:fl-dis}
	M\ddot{x} + \Gamma\dot{x} = -\frac{\rmd U(x)}{\rmd{x}} + A\cos(\Omega t) + F + \sqrt{2\Gamma k_BT}\, \xi(t),
\end{equation}
where dot means differentiation with respect to time $t$ and $x$ is the position of the particle. The factor $\sqrt{2\Gamma k_BT}$ follows from the fluctuation-dissipation theorem \cite{marconi2008} and ensures the correct Gibbs equilibrium state for the free particle.
The potential, the harmonic force and thermal fluctuations are symmetric with respect to time and space, so the only perturbation that breaks the symmetry of Eq. \eref{eq:fl-dis} and allows for the emergence of directed transport is the constant force $F$.

There are many physical systems that can be modeled in terms of the dynamics given by Eq. (\ref{eq:fl-dis}) including superionic conductors \cite{fulde1975,dieterich1980}, dipoles rotating in external fields \cite{langevin}, charge density waves \cite{gruner1981}, Josephson junctions \cite{kautz,blackburn2016} and its variations like SQUIDs \cite{spiechowicz2015chaos, spiechowicz2019chaos} as well as cold atoms dwelling in optical lattices \cite{lutz2013, lutz2017}, to name but a few.

In physics only the relations between characteristic scales of time, length and energy are relevant for the progress of observed phenomena, but not their absolute values. This fact suggests recasting of Eq. \eref{eq:fl-dis} into its dimensionless form in which all quantities are expressed as combinations of characteristic scales for the system. This procedure makes the analysis independent of the experimental setup provided that the mathematical structure of Eq. \eref{eq:fl-dis} is preserved. It also allows to reduce the number of free parameters appearing in the model. The scaling is based on choosing the length and time scales. The most obvious selection for the characteristic length is the potential period $L$. Depending on the definition of the time scale one arrives at different scalings. Below two most common variants are presented and the differences between them are elaborated.
\subsection{First scaling with mass $m=1$}
The first time scale follows from the equation for frictionless movement of a particle in the periodic potential
\begin{equation}
	M\ddot{x} = -U'(x).
\end{equation}
In such a case the time unit can be extracted as follows \cite{hanggi2020}
\begin{equation}
	\tau_\vgamma = L\sqrt{\frac{M}{\Delta U}}.
\end{equation}
It is related to the period of linearized oscillations within one potential well. We define the dimensionless particle coordinate and time as
\begin{equation}
	\hat{x} = \frac{x}{L}, \quad t_1 = \frac{t}{\tau_1}.
\end{equation}
Under such a choice the original Eq. \eref{eq:fl-dis} takes the following form
\begin{equation} \label{eq:scl-vgamma}
	\dot{v}_\vgamma + \gamma v_\vgamma = -\frac{\rmd \hat{U}(\hat{x})}{\rmd \hat{x}} + a\cos(\omega_\vgamma t_\vgamma) + f + \sqrt{2\gamma D}\, \hat{\xi} (t_\vgamma),
\end{equation}
where $v_\vgamma = \rmd \hat{x} / \rmd t_\vgamma$. The rescaled potential is defined as
\begin{equation}
	\hat{U}(\hat{x}) = \frac{U(L\hat{x})}{\Delta U} = \sin(2\pi\hat{x})
\end{equation}
and possesses the period equal to unity. The dimensionless thermal noise reads
\begin{equation}
	\hat{\xi}(t_1) = \frac{L}{\Delta U} \xi(t) = \frac{L}{\Delta U} \xi(\tau_1 t_1)
\end{equation}
and still represents the $\delta$-correlated Gaussian white noise of zero mean, c.f. Eq. (2). Its rescaled intensity $D$ is a ratio of the thermal and potential barrier energies 
\begin{equation}
	D = \frac{k_BT}{\Delta U}.
\end{equation}
The external force parameters read
\begin{equation}
	a = \frac{L}{\Delta U}A,\quad
	f = \frac{L}{\Delta U}F.
\end{equation}
The reader can note that in this scaling one may introduce also the dimensionless mass but it is fixed to $m=1$. The rest of quantities appearing in Eq. \eref{eq:scl-vgamma} explicitly depend on the chosen time unit, namely
\begin{equation}
	\gamma = \frac{\tau_1}{\tau_0} = \frac{\tau_1}{M}\Gamma = \frac{L}{\sqrt{M\Delta U}}\Gamma,\quad
	\omega_\vgamma = \tau_\vgamma\Omega.
\end{equation}
The dimensionless friction coefficient $\gamma$ can be expressed as a ratio of two characteristic time scales where $\tau_0 = M/\Gamma$ stands for the so-called Langevin time, i.e. the relaxation time for the velocity of a free Brownian particle. Nevertheless it is instructive to note that $\gamma$ is proportional to the actual physical friction coefficient $\Gamma$.
\subsection{Second scaling with friction coefficient $\gamma=1$}
Another time scale can be extracted from the equation of an overdamped motion of a particle in the periodic potential
\begin{equation}
	\Gamma \dot{x} = -U'(x).
\end{equation}
The time unit is then \cite{hanggi2020}
\begin{equation}
	\tau_\vm = \frac{\Gamma L^2}{\Delta U}.
\end{equation}
It scales the characteristic time for the overdamped particle to move from the maximum to minimum of the potential $U(x)$. In this case the original Eq. \eref{eq:fl-dis} transform as follows
\begin{equation} \label{eq:scl-vm}
	m\dot{v}_\vm + v_\vm = -\frac{\rmd \hat{U}(\hat{x})}{\rmd \hat{x}} + a\cos(\omega_\vm t_\vm) + f + \sqrt{2D}\, \hat{\xi}(t_\vm),
\end{equation}
where now $v_\vm = \rmd \hat{x} / \rmd t_\vm$.
The parameters that do not depend on the time scale are the same as in the first scaling.
The new quantities are
\begin{equation}
	t_\vm = \frac{t}{\tau_\vm},\quad
	m = \frac{\tau_0}{\tau_2} = \frac{1}{\Gamma \tau_\vm}M = \frac{\Delta U}{\Gamma^2 L^2}M,\quad
	\omega_\vm = \tau_\vm\Omega.
\end{equation}
Now the dimensionless mass $m$ is proportional to the physical mass $M$ and expressed as a ratio of two characteristic time scales $\tau_0 = M/\Gamma$ and $\tau_2$ while the friction coefficient formally scales to $\gamma=1$.
\subsection{Differences between scalings}
Since the two above presented variants of the rescaled dynamics refer to the same model given by Eq. \eref{eq:fl-dis} the parameter definitions appearing in both of them must be associated with each other. One can show that they obey the following relations
\begin{equation} \label{eq:scl:gamma-m}
	t_\vgamma = \frac{1}{\sqrt{m}}t_\vm,\quad \gamma = \frac{1}{\sqrt{m}},\quad \omega_\vgamma = \sqrt{m}\omega_\vm.
\end{equation}
Obviously, the inverse is also true
\begin{equation} \label{eq:scl:m-gamma}
	t_\vm = \frac{1}{\gamma}t_\vgamma,\quad m = \frac{1}{\gamma^2},\quad \omega_\vm = \gamma\omega_\vgamma.
\end{equation}
It is important to note that the velocities $v_\vgamma$ and $v_\vm$ are calculated as derivatives of $\hat{x}$ with respect to $t_\vgamma$ and $t_\vm$, respectively.
Thus, they cannot be compared directly, but a proper rescaling is needed
\begin{equation}
	v_\vgamma = \sqrt{m}v_\vm,\quad v_\vm = \gamma v_\vgamma.
\end{equation}
As long as fixed values of $\gamma$ and $m$ are considered both scalings are equivalent. One can always recalculate all quantities from one scaling to the other. It might seem that thermal fluctuations terms are not equivalent since in the first scaling the thermal noise prefactor depends on both $\gamma$ and $D$, whereas in the second one it is determined only by $D$. However, since the white noise is $\delta$-correlated, one can show that
\begin{equation}
	\sqrt{2D}\,\hat{\xi}(t_\vm) = \sqrt{2D}\,\hat{\xi}(t_\vgamma/\gamma) = \sqrt{2 \gamma D}\,\hat{\xi}(t_\vgamma),
\end{equation}
which means that these terms are also equivalent.

Nevertheless, there is one very important difference between these scalings which is the relation of the dimensionless time $t_1$ and $t_2$ to the actual physical time $t$ in \mbox{Eq. \eref{eq:fl-dis}}. In the first scaling the time unit $\tau_\vgamma$ does not depend on the friction coefficient $\Gamma$, but on the mass $M$. It means that one can easily investigate the impact of damping $\Gamma$ in this scaling without changing the time scale. Analyzing the influence of mass $M$ is also possible, but for every value of $M$ the time scale in the system would be different. %This also means that the velocities for different values of $M$ could not be compared. 
Interpretation of the results in such an approach would be hardly possible. Similarly, the time unit $\tau_\vm$ in the second scaling does not depend on the mass $M$, but  on the damping $\Gamma$, which makes it unsuitable for studying the impact of friction $\Gamma$ on the dynamics.

This difference has the most profound consequences in the limiting situations when either $\gamma$ or $m$ approaches zero which is the case in the Hamiltonian and overdamped dynamical regimes, respectively. When $\gamma \to 0$, then from the relation $t_\vm = t_\vgamma/\gamma$ it follows that $t_\vm \to \infty$. This means that taking the limit $\gamma \to 0$ in the first scaling is \emph{not equivalent} to requiring $m \to \infty$ in the second one. %since the corresponding times could not be compared.
Similarly, the relation $v_\vm = \gamma v_\vgamma$ implies that the velocity in the second scaling would then tend to zero. For these reasons, the approach to the case of Hamiltonian dynamics in which formally $\gamma = 0$ have to be analyzed by taking the limit $\gamma \to 0$ in the first scaling. The analogous discussion could be repeated for the overdamped regime when $m = 0$ that would lead us to the conclusion that such a scenario must be investigated by performing the limit $m \to 0$ in the second scaling what is \emph{not equivalent} to the case $\gamma \to \infty$ in the first one.
%
%Moreover, the proportionality relation $\gamma \propto \Gamma$ between the dimensionless and physical friction coefficient makes the first scaling much more intuitive in analyzing the impact of damping on the system dynamics. Similarly, the dimensionless mass $m \propto M$ and therefore the second scaling is more convenient for investigating the influence of inertia on the particle kinetics.

In the remaining part of the article we will use only the dimensionless variables and therefore for simplicity we omit the hat-notation, i.e. we will write $x$ instead of $\hat{x}$ and so on.
\subsection{Quantity of interest}
The most fundamental quantity characterizing the directed transport is the average velocity defined as
\begin{equation} \label{eq:vavg}
	\langle v \rangle = \lim\limits_{t\to\infty} \frac{1}{t} \int_{0}^{t} ds \langle \dot{x}(s) \rangle,
\end{equation}
where $\langle \cdot \rangle$ indicates averaging over all thermal noise realizations as well as initial conditions for the particle position $x(0)$ and velocity $v(0)$.
%where $\langle \cdot \rangle$ indicates averaging over all thermal noise realizations, initial conditions for the particle position $x(0)$ and velocity $v(0)$ as well as the phase of the harmonic driving $a\cos(\omega t)$. 
The latter is obligatory especially in the limiting case of deterministic dynamics $D = 0$ when the ergodicity of the system may be broken and consequently the results are affected by those initial conditions \cite{spiechowicz2016scirep, spiechowicz2021pre2}. The above definition is independent on the scaling selection. The time $t$ can be either $t_\vgamma$ or $t_\vm$ and the velocity $v$ may be $v_\vgamma$ or $v_\vm$.
\section{Absolute negative mobility effect} \label{sec:negative_mobility}
To make the paper self-contained in this part we briefly review the state of the art of the absolute negative mobility effect for the studied system.

We start with the observation that the underlying symmetries of the Langevin \mbox{Eq. \eref{eq:fl-dis}} imply that the average velocity $\langle v \rangle$ is an odd function of the constant bias $f$, namely $\langle v \rangle(-f) = -\langle v \rangle(-f)$ \cite{denisov2014}. Therefore the directed transport cannot emerge in absence of the static force $f$ as then $\langle v \rangle (0) \equiv 0$. We define the mobility $\mu$ \cite{kostur2008} of the particle as
\begin{equation}
	\langle v \rangle(f) = \mu(f)f
\end{equation}
to describe its ability to move through the medium in response to the acting force. Typically the resultant particle displacement follows the direction of the applied bias $\mu(f) > 0$. The corresponding average velocity $\langle v \rangle$ renders a nonlinear function of the constant force $f$ and it is expected that $\langle v \rangle$ increases for growing $f$. In the linear response regime the velocity $\langle v \rangle = \mu_0 f$ is a linear function of the force $f$ with the constant mobility coefficient $\mu_0$ \cite{kostur2008}. 

The term absolute negative mobility refers to the paradoxical case when the net particle movement is opposite to the direction of the static load around zero bias \cite{slapik2019prl}, i.e.
\begin{equation}
	\mbox{for} \quad f > 0: \quad \langle v \rangle = \mu(f) f < 0 \quad \Longrightarrow \quad \mu(f) < 0.
\end{equation}

The counterintuitiveness of this phenomenon follows from the fact that in a linear system the influence of all forces can be analyzed separately and the collective effect is simply a sum of responses induced by all perturbations. Since the harmonic driving $a\cos{(\omega t)}$ and thermal fluctuations $\xi(t)$ have a vanishing mean, in a linear system their contribution to the net movement would be zero and the particle would follow the direction of the constant force $f$, what implies that the absolute negative mobility effect would not emerge. However, the potential $U(x)$ is periodic and hence gives rise to a nonlinear system, where the superposition principle is no longer valid. The nonlinearity is another necessary condition for $\mu(f) < 0$ to occur \cite{machura2007,speer2007}.

One can argue that the net movement of the particle in a direction opposite to the constant force contradicts the Le Chatelier-Braun's principle \cite{machura2007,speer2007}. This law is however no longer valid for systems out of equilibrium. Therefore the key requirement for the occurrence of the absolute negative mobility is that the system is driven far from thermal equilibrium into a nonequilibrium state \cite{machura2007,speer2007}. In the considered model it is guaranteed by the presence of the external time periodic driving $a\cos{(\omega t)}$.

%The last condition that determines the occurrence of the negative directed transport is friction.
Last but not least, it has been already shown in literature that the absolute negative mobility does not emerge in the limiting case of one-dimensional overdamped ($m = 0$) and Hamiltonian ($\gamma=0$) regimes \cite{slapik2018}. Omitting the dissipative term $-\gamma v$ is equivalent to the situation in which the system is coupled to infinitely hot bath for which the potential $U(x)$ term becomes negligible. As it was discussed above the latter is essential for the emergence of the absolute negative mobility effect.

Three different mechanisms responsible for this counterintuitive phenomenon are currently known -- deterministic chaotic, deterministic non-chaotic and thermal noise induced \cite{machura2007,speer2007,slapik2018}. The deterministic dynamics given by Eq. \eref{eq:scl-vgamma} or Eq. \eref{eq:scl-vm} with $D = 0$ can be recasted into a set of three autonomous differential equations of the first order for which the corresponding phase space is three-dimensional $\{x, v, z = \omega t\}$ being the minimal requirement for the system to display chaotic evolution \cite{strogatz}. In such a case the absolute negative mobility effect emerges as a result of the subtle interplay between coexisting attractors and transient chaos \cite{speer2007}. Recently it has been demonstrated that this phenomenon can occur also in the deterministic system given by Eq. \eref{eq:scl-vm} which is in the non-chaotic dynamical regime \cite{slapik2018}, i.e. it exhibits regular attractors transporting the particle in the direction opposite $\langle v \rangle < 0$ to the applied bias $f > 0$. Finally, the absolute negative mobility can be triggered solely by thermal equilibrium fluctuations \cite{machura2007}. In such a case in the deterministic dynamics the absolute mobility of the particle is positive $\mu(f) > 0$ but it takes negative values $\mu(f) < 0$ for certain temperature $D$ window.

\section{Methods}
The Fokker-Planck equation corresponding to Eq. \eref{eq:scl-vgamma} or Eq. \eref{eq:scl-vm} is the second order parabolic partial differential equation with a nonlinear and time periodic drift coefficient due to the presence of potential $U(x)$ and driving $a\cos{(\omega t)}$, respectively. For this reason its solution is unattainable analytically and in order to analyze the transport behaviour of driven Brownian particle we carried out comprehensive numerical simulations. The studied system possesses a complex five dimensional parameter space $\{\gamma\, \mbox{or}\, m, a, \omega, f, D\}$. Its systematic exploration was not possible until very recently due to limited computational capabilities of modern hardware and lack of innovative implementations of simulation methods. We performed numerical analysis by harvesting the GPU supercomputers \cite{spiechowicz2015cpc} that allowed us to draw both qualitative and quantitative conclusions about the emergence of absolute negative mobility phenomenon in the parameter space. The latter were picked from a cuboid in $\{\gamma, a, \omega\}$ or $\{m, a, \omega\}$ sub-space containing $400\times400\times400$ values. This volume, with 14 combinations of $f$ and $D$ values, resulted in nearly $10^9$ parameter regimes per each considered scaling.

We employed a weak second order predictor-corrector scheme \cite{platen} to simulate stochastic dynamics given by Eq. \eref{eq:scl-vgamma} or Eq. \eref{eq:scl-vm}. The time step of integration was scaled as $h = 10^{-2} \times \mathsf{T}$ where $\mathsf{T} = 2\pi/\omega$ is the fundamental period of the external driving $a\cos{(\omega t)}$. The average velocity $\langle v \rangle$ was calculated over the ensemble of $1024$ system trajectories each starting with different initial conditions. The initial positions $x(0)$ and velocities $v(0)$ were uniformly distributed over the intervals $[0,1]$ and $[-2,2]$, respectively. All trajectories lasted for $10^3$ periods $\mathsf{T}$ of the external driving and spanned the interval used for calculating the time average in the definition of directed transport $\langle v \rangle$.

Since the latter quantity is invariant under changes of the sign of $a$, we restrict our analysis only to positive values $a\in[0, 25]$. The limit can be put also on the frequency $\omega$. For $\omega \to 0$ the adiabatic approximation may be employed and the velocity follows the external force. On the other hand, for $\omega \to \infty$ the average velocity can be expressed via the Bessel functions \cite{kautz} and the absolute negative mobility does not emerge. Consequently, in our simulations we analyzed the interval $\omega \in [0.01, 20]$. Moreover, since the directed transport $\langle v \rangle$ is an odd function of the static bias $f$, it is sufficient to consider only positive values $f > 0$ of the latter parameter. It is intuitive that when the constant force $f$ is much larger than other perturbations it dominates the dynamics and the Brownian particle velocity follows its direction. Therefore the values of $f$ were chosen from the interval $f \in [0.01,1.5]$. Finally, we remind that the absolute negative mobility does not occur in the overdamped $m = 0$ and Hamiltonian $\gamma = 0$ regimes. Consequently, the simulations were performed for $\gamma \in [0.1, 10]$ and $m \in [0.01, 10]$. In majority of the investigated parameter sets the thermal noise had a destructive influence on the occurrence of absolute negative mobility. For this reason most of the regimes corresponded to the deterministic system with $D=0$, with a few runs for temperature up to $D = 0.01$. Overall, the analyzed parameter subspace allowed to cover almost entire range of values for which the net movement of the particle is in the direction opposite to the applied force and the absolute negative mobility phenomenon emerges.
\section{Results}
Although the simulations were performed for a wide range of parameters, the below presented results show only some subsets of the studied region. %The areas where the directed transport did not occur or was only positive were cut off. In some cases the negative mobility could be seen for a broad span of the parameter values, however to keep the clarity and reveal the details, only the most interesting parts were shown. 
The presented areas were chosen to reflect the general dependence of the directed transport $\langle v \rangle$ on the model parameters.
\begin{figure}[t]
	\centering
	\includegraphics[width=1.0\linewidth]{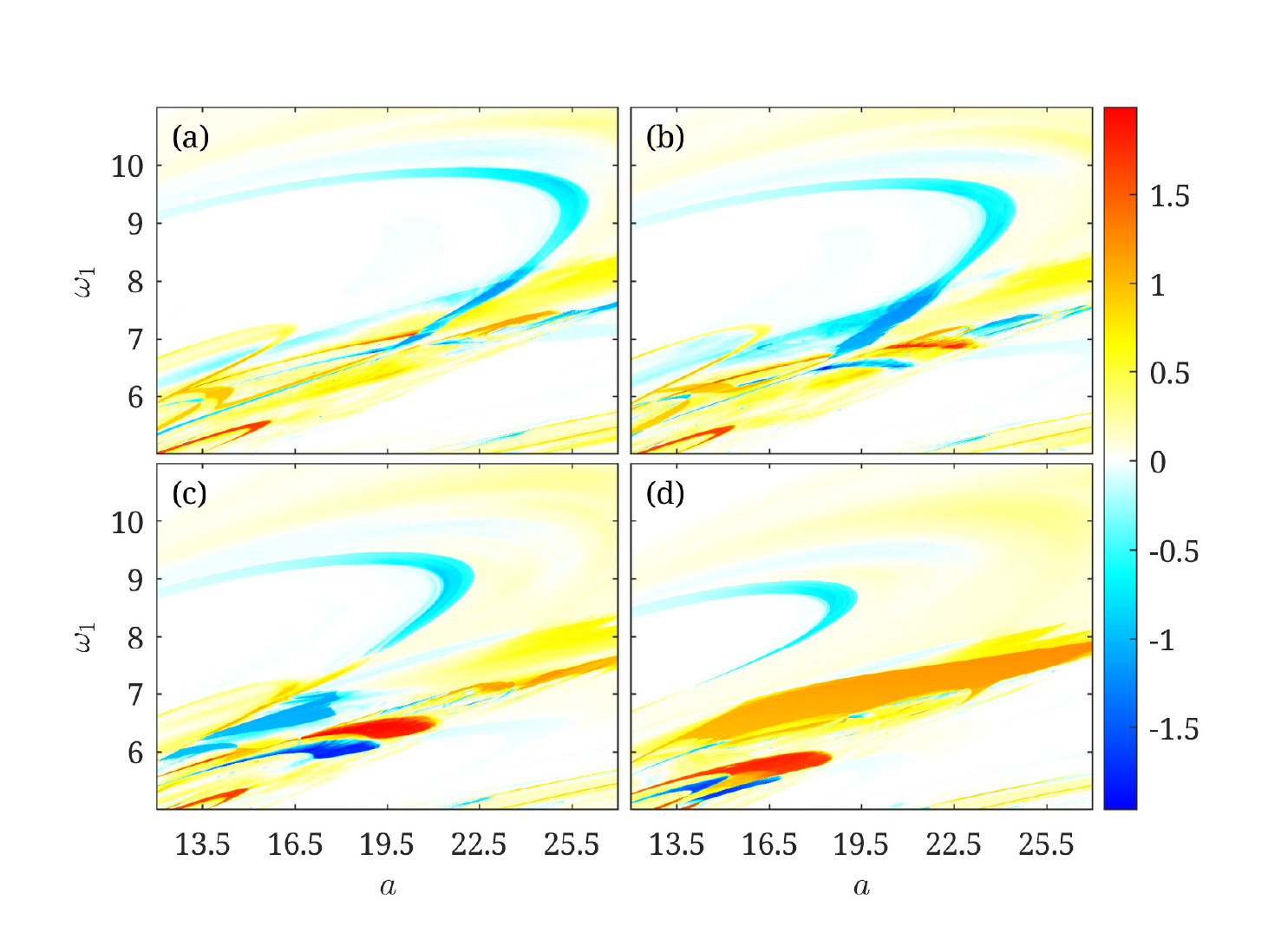}
	\caption{Influence of dissipation $\gamma$ on the absolute negative mobility phenomenon. The average velocity $\langle v_\vgamma \rangle$ of the Brownian particle versus amplitude $a$ and frequency $\omega_\vgamma$ of the external driving for different values of $\gamma$ with $f=0.1$ and $D=0$. Panel (a)~$\gamma=1.0839$, (b)~$\gamma=1.1749$, (c)~$\gamma=1.3032$, (d)~$\gamma=1.4962$.}
	\label{fig:gaw:g}
\end{figure}

\subsection{First scaling with mass $m=1$}
We start our analysis with the investigation of the absolute negative mobility effect in the subspace of parameters characterizing the external harmonic driving. In Fig. \ref{fig:gaw:g} we depict the influence of dissipation $\gamma$ on the absolute negative mobility phenomenon. The average velocity $\langle v_\vgamma \rangle$ of the Brownian particle as a function of amplitude $a$ and angular frequency $\omega_\vgamma$ of the harmonic driving for four different values of $\gamma$ with $f=0.1$ and $D=0$ is presented.
One can observe an arc-like area of absolute negative mobility and its evolution with the change of $\gamma$.
\emph{When $\gamma$ increases, this region moves towards lower $a$ and $\omega_\vgamma$.}
For $\gamma=1.1749$ this area is maximal. When $\gamma$ increases further the whole structure vanishes and adjacent regions of positive velocity become more intense. There is also a small area of relatively low negative velocity below the arc, accompanied by a similar region of positive velocity.
%The negative area is the largest for $\gamma=1.3032$, but it is visible for only a small range of values of $\gamma$.
\begin{figure}[t]
	\centering
	\includegraphics[width=1.0\linewidth]{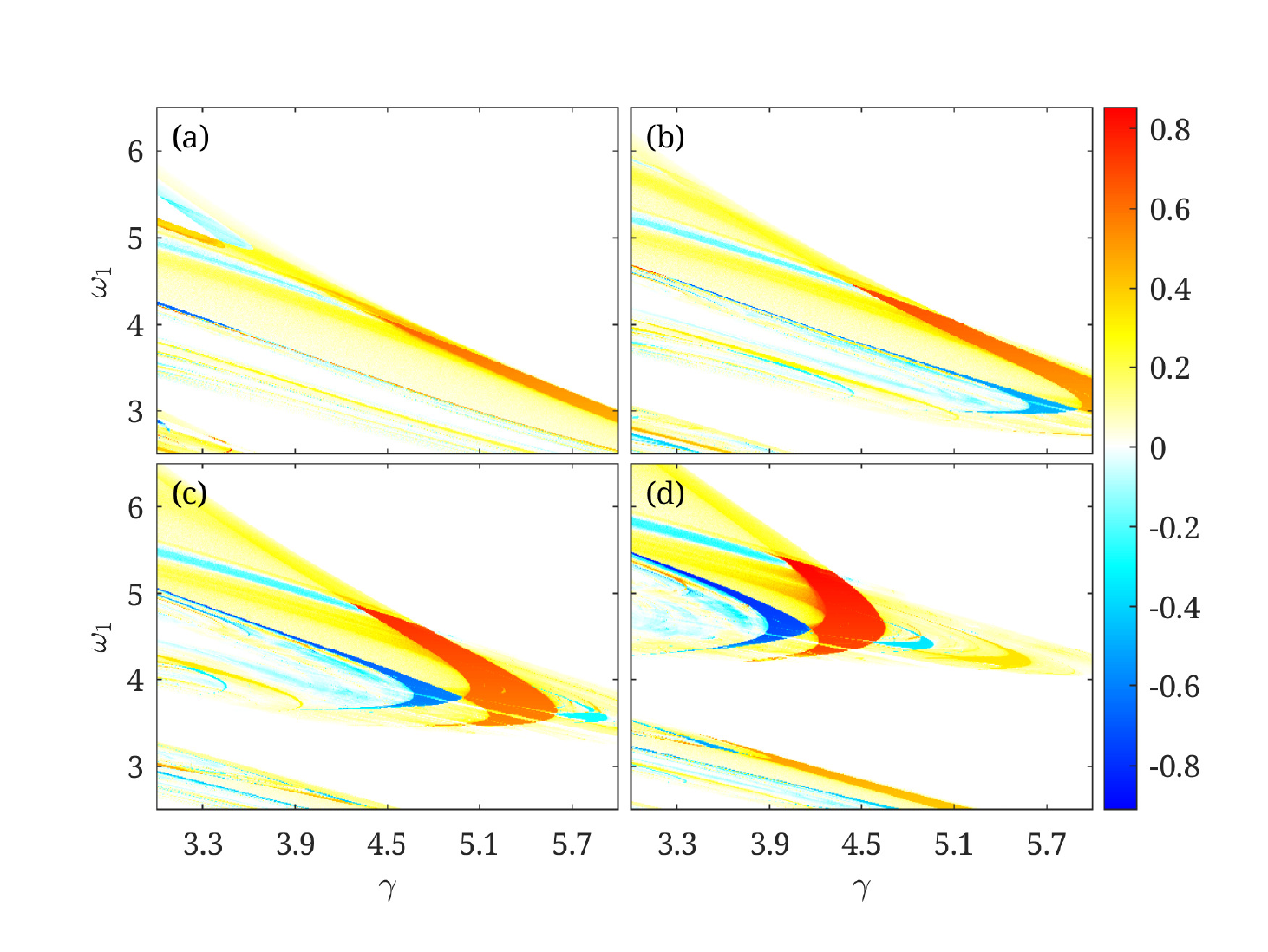}
	\caption{Impact of the amplitude $a$ of the external driving on the absolute negative mobility effect. The average velocity $\langle v_\vgamma \rangle$ of the Brownian particle versus friction coefficient $\gamma$ and the frequency $\omega_\vgamma$ of the driving for different values of $a$ with $f=0.1$ and $D=0$. Panel (a)~$a=9.25$, (b)~$a=10.375$, (c)~$a=11.5625$, (d)~$a=13.25$.}
	\label{fig:gaw:a}
\end{figure}

Fig. \ref{fig:gaw:a} presents the impact of the amplitude $a$ of the external harmonic driving on the absolute negative mobility effect. This panel depicts the average velocity $\langle v_\vgamma \rangle$ of the Brownian particle as a function of friction coefficient $\gamma$ and angular frequency $\omega_\vgamma$ of the harmonic driving for different values of $a$ with $f=0.1$ and $D=0$.
There are two main areas of V-like shape, one with negative, the other with positive velocity.
%Most likely they represent two similar mechanisms that induce transport in the system.
In the places where they overlap, they compensate each other and the velocity in these overlapping regions is much closer to zero than in the non-overlapping parts.
\emph{When $a$ grows the structure moves towards smaller $\gamma$ and greater $\omega_\vgamma$.}
Moreover, for increasing $a$ the velocity in both areas assumes more extreme values.
%One can also see similar positive and negative structures for higher values of $\gamma$.
%They also coexist in a pair, but they are smaller and absolute values of the velocity in these areas are lower than in the bigger ones.
\begin{figure}[t]
	\centering
	\includegraphics[width=1.0\linewidth]{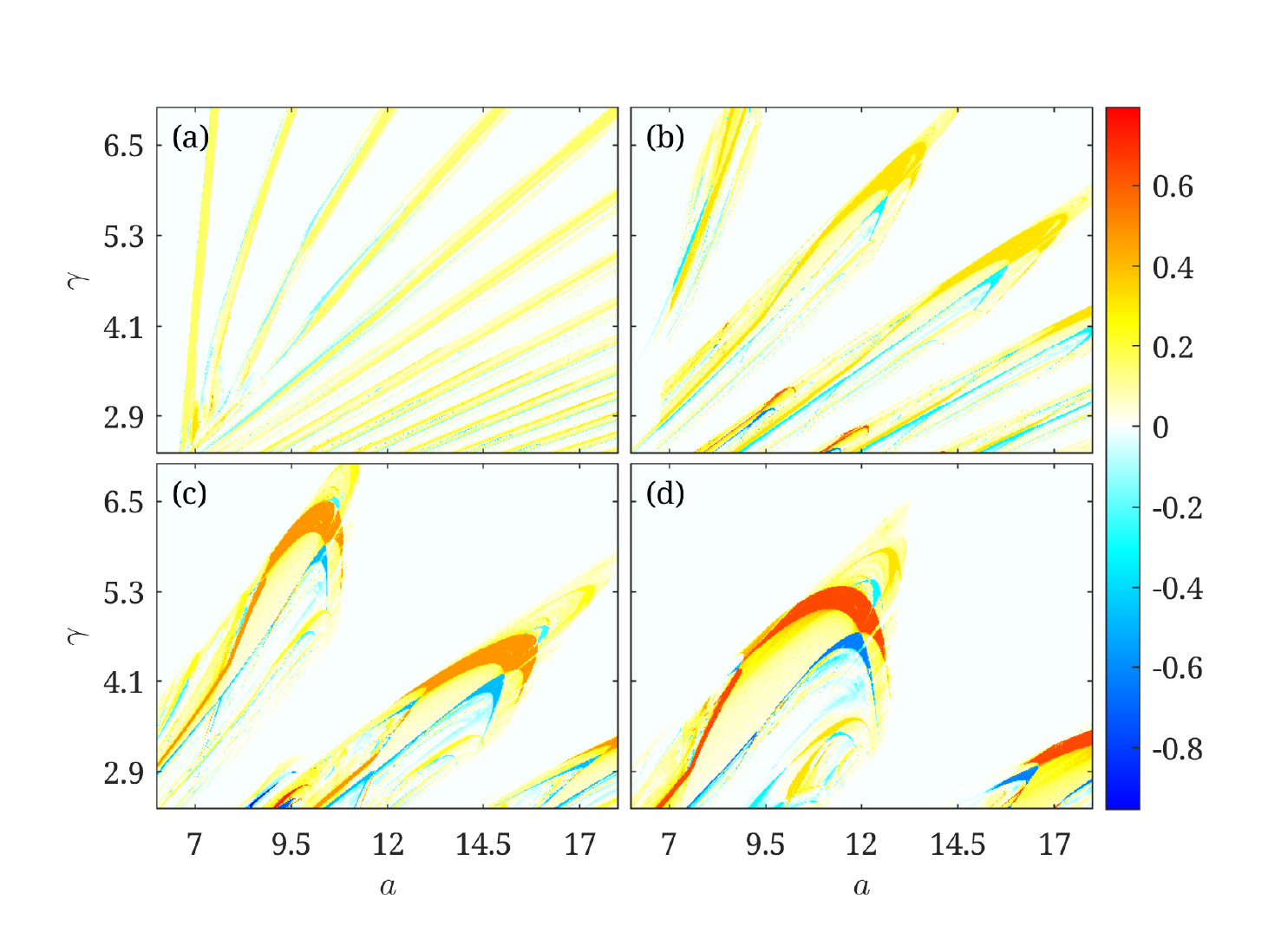}
	\caption{Response of the absolute negative mobility induced by a change of the external driving frequency $\omega_1$. The average velocity $\langle v_\vgamma \rangle$ of the Brownian particle versus amplitude $a$ of the driving and friction coefficient $\gamma$ for different values of $\omega_\vgamma$ with $f=0.1$ and $D=0$. Panel (a)~$\omega_\vgamma=1.0$, (b)~$\omega_\vgamma=2.0$, (c)~$\omega_\vgamma=3.0$, (d)~$\omega_\vgamma=4.0$.}
	\label{fig:gaw:w}
\end{figure}

In Fig. \ref{fig:gaw:w} we illustrate the response of the absolute negative mobility effect induced by a change of the external driving frequency $\omega_1$. The directed transport $\langle v_\vgamma \rangle$ of the Brownian particle versus the amplitude $a$ and friction coefficient $\gamma$ for selected values of $\omega_\vgamma$ and fixed $f=0.1$ and $D=0$ is presented there.
For small $\omega_\vgamma$ there are ray-like structures with positive velocity with very thin stripes of the absolute negative mobility along them.
It suggests that for small $\omega_\vgamma$ there is almost linear relation between $a$ and $\gamma$ for which the transport occur. % proportional + constant
The ray-like pattern is deformed if $\omega_\vgamma$ grows. First, the areas become larger and the absolute value of the average velocity increases. Second, they tend to orientate more horizontally indicating that \emph{if $\omega_\vgamma$ is increased the absolute negative mobility is expected to emerge for greater $a$ but smaller $\gamma$.} At some point the ray-like structure is changed to V-like, similar to the one visible on Fig. \ref{fig:gaw:a}. The areas of positive and negative velocity occur in pairs and partially overlap.
\begin{figure}[t]
	\centering
	\includegraphics[width=1.0\linewidth]{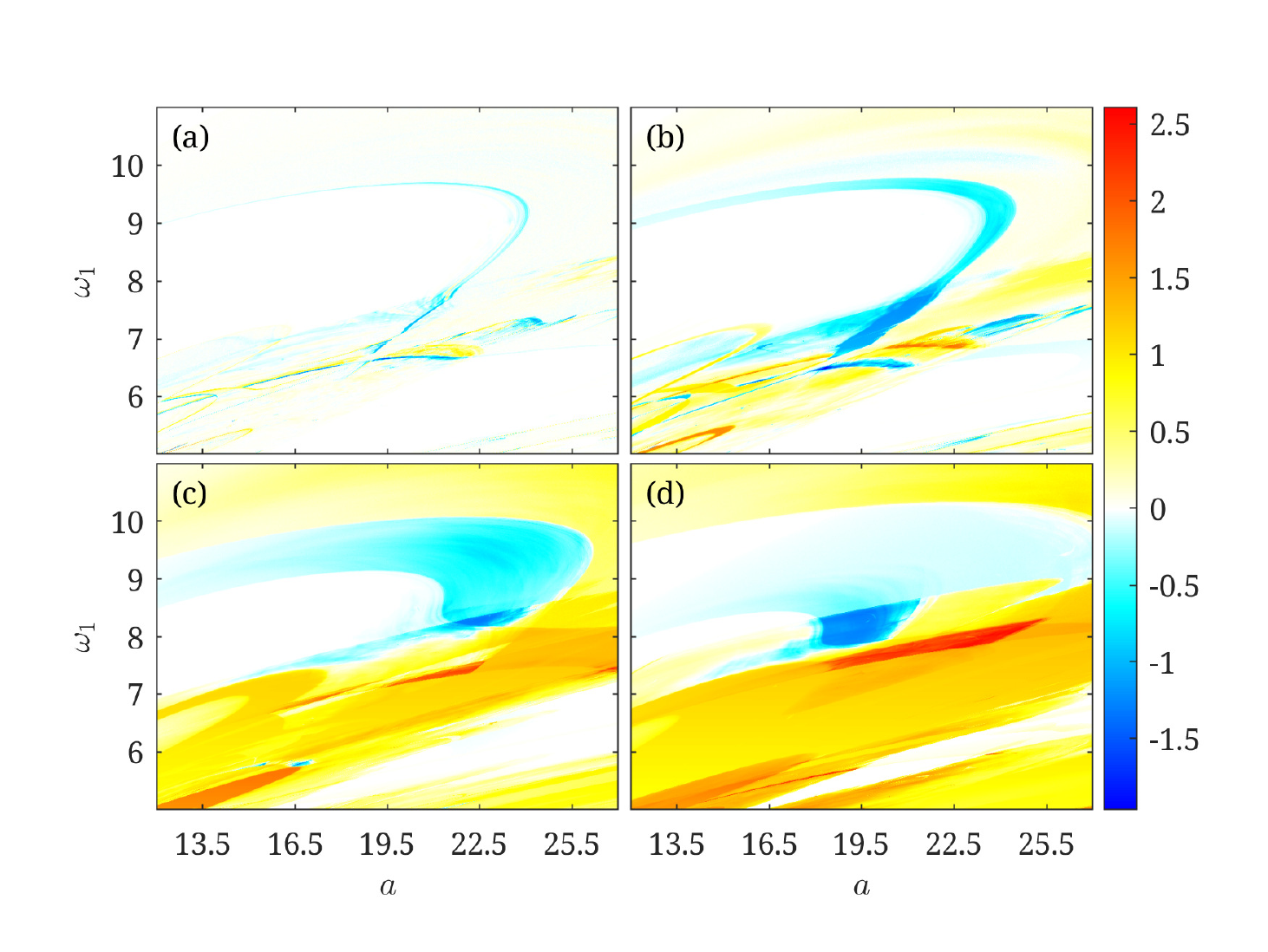}
	\caption{Impact of the static bias $f$ on the absolute negative mobility. The average velocity $\langle v_\vgamma \rangle$ of the Brownian particle versus amplitude $a$ of the driving and friction coefficient $\gamma$ for different values of $f$ with $\gamma=1.1749$ and $D=0$. Panel (a)~$f=0.01$, (b)~$f=0.1$, (c)~$f=0.5$, (d)~$f=1.0$.}
	\label{fig:gaw:f}
\end{figure}

Next, Fig. \ref{fig:gaw:f} shows the impact of the static bias $f$ on the absolute negative mobility. The panel presents the average velocity $\langle v_\vgamma \rangle$ of the Brownian particle as a function of the amplitude $a$ and the frequency $\omega_\vgamma$ of the harmonic driving for different $f$ with $\gamma=1.1749$ and $D=0$.
For small $f$ the absolute negative mobility areas are barely visible, similarly to the regions of positive ones.
This is consistent with the already mentioned requirement which states that the external constant force $f$ breaks the symmetry of the system and is necessary to induce the transport.
When $f$ increases, the region of absolute negative mobility broadens, however for large $f$ the absolute value of negative velocities decreases and the structure finally vanishes.
%For higher $f$ the surrounding areas of positive velocity also increase and finally cover the whole parameter space.
The reason is that when the static bias $f$ is much larger than other forces in the system, it dominates the dynamics and the absolute negative mobility effect cannot occur.
The region of this anomalous transport behaviour is largest for $f=0.5$, however for $f=0.1$ there are many finer structures that vanish for greater $f$.
\begin{figure}[t]
	\centering
	\includegraphics[width=1.0\linewidth]{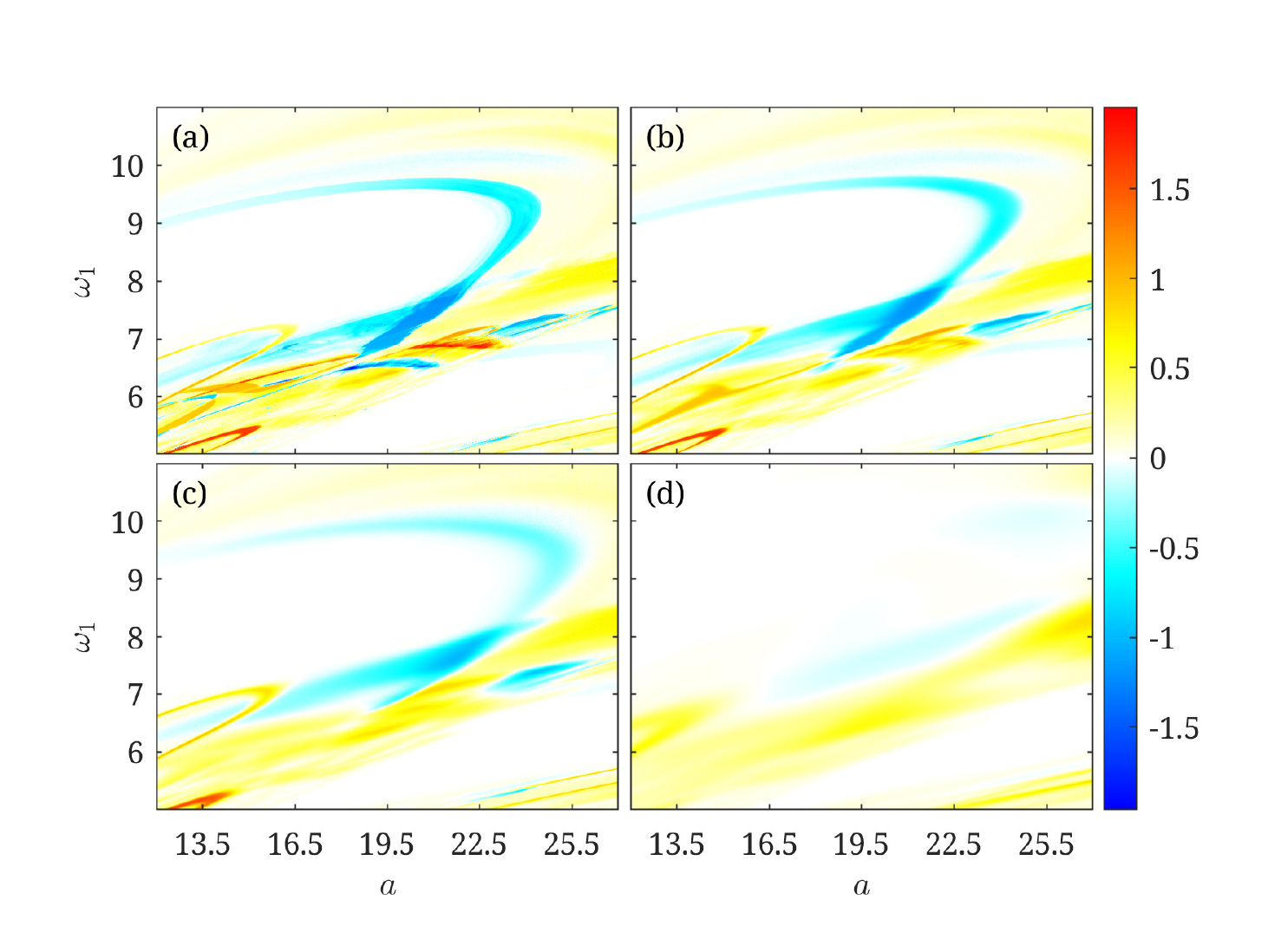}
	\caption{Influence of temperature $D$ on the absolute negative mobility. The average velocity $\langle v_\vgamma \rangle$ of the Brownian particle versus amplitude $a$ of the harmonic driving and friction coefficient $\gamma$ is depicted for different values of $D$ with $\gamma=1.1749$ and $f=0.1$. Panel (a)~$D=0$, (b)~$D=0.0001$, (c)~$D=0.001$, (d)~$D=0.01$.}
	\label{fig:gaw:D}
\end{figure}

Finally, in Fig. \ref{fig:gaw:D} we discuss the influence of temperature $D$ on the absolute negative mobility. The average velocity $\langle v_\vgamma \rangle$ of the Brownian particle versus the amplitude $a$ of the harmonic driving and dissipation $\gamma$ is presented there for various values of $D$ with $\gamma = 1.1749$ and $f = 0.1$. The reader can observe that temperature growth causes blurring of the structures of both positive and negative velocity. The fine details disappear first while the larger one are still present in the noisy case, however, for high enough temperature they all vanish. It is expected as eventually thermal noise dominates the dynamics and the particle behaves as the free one. \emph{Typically temperature has destructive impact on the emergence of absolute negative mobility in the parameter space}.
\begin{figure}[t]
	\centering
	\includegraphics[width=1.0\linewidth]{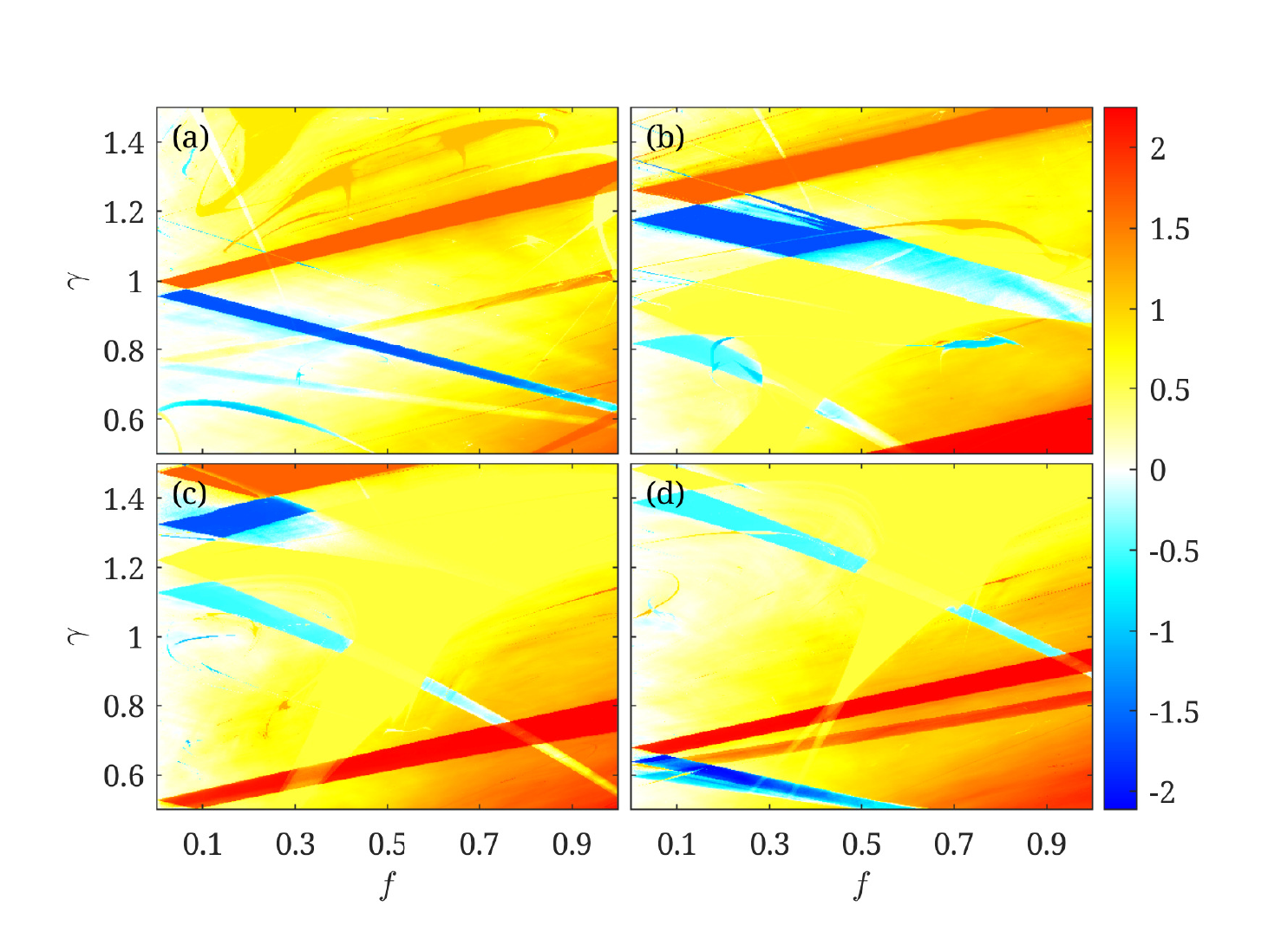}
	\caption{The average velocity $\langle v_\vgamma \rangle$ of the Brownian particle as a function of the constant force $f$ and friction coefficient $\gamma$ for different values of external driving amplitude $a$ with $\omega_\vgamma=3.5$ and $D=0$. Panel (a)~$a=9.25$, (b)~$a=10.0$, (c)~$a=10.5$, (d)~$a=11.0$.}
	\label{fig:gfw:a}
\end{figure}

Now we turn to the analysis of different parameter subspace associated with the propelling force $f$ and dissipation $\gamma$. In Fig. \ref{fig:gfw:a} we present the average velocity $\langle v_1 \rangle$ of the Brownian particle versus the static bias $f$ and damping $\gamma$ for different values of the external driving amplitude $a$ with $\omega_1 = 3.5$ and $D = 0$. In all plots pairs of wedge-like areas with positive and negative velocity are visible. In places where they overlap they compensate each other. The borders of these regions are approximately linear, which suggest a simple relation between $f$ and $\gamma$ for which the directed transport occurs. \emph{The area of anomalous transport behaviour moves towards lower values of $\gamma$ as $f$ is increased.} %For higher values of $a$ the structures move toward greater values of $\gamma$ and broaden.
%Moreover, The reader can observe that \emph{if the amplitude $a$ grows the negative mobility phenomenon occurs for greater values of dissipation $\gamma$}.
 
\begin{figure}[t]
	\centering
	\includegraphics[width=1.0\linewidth]{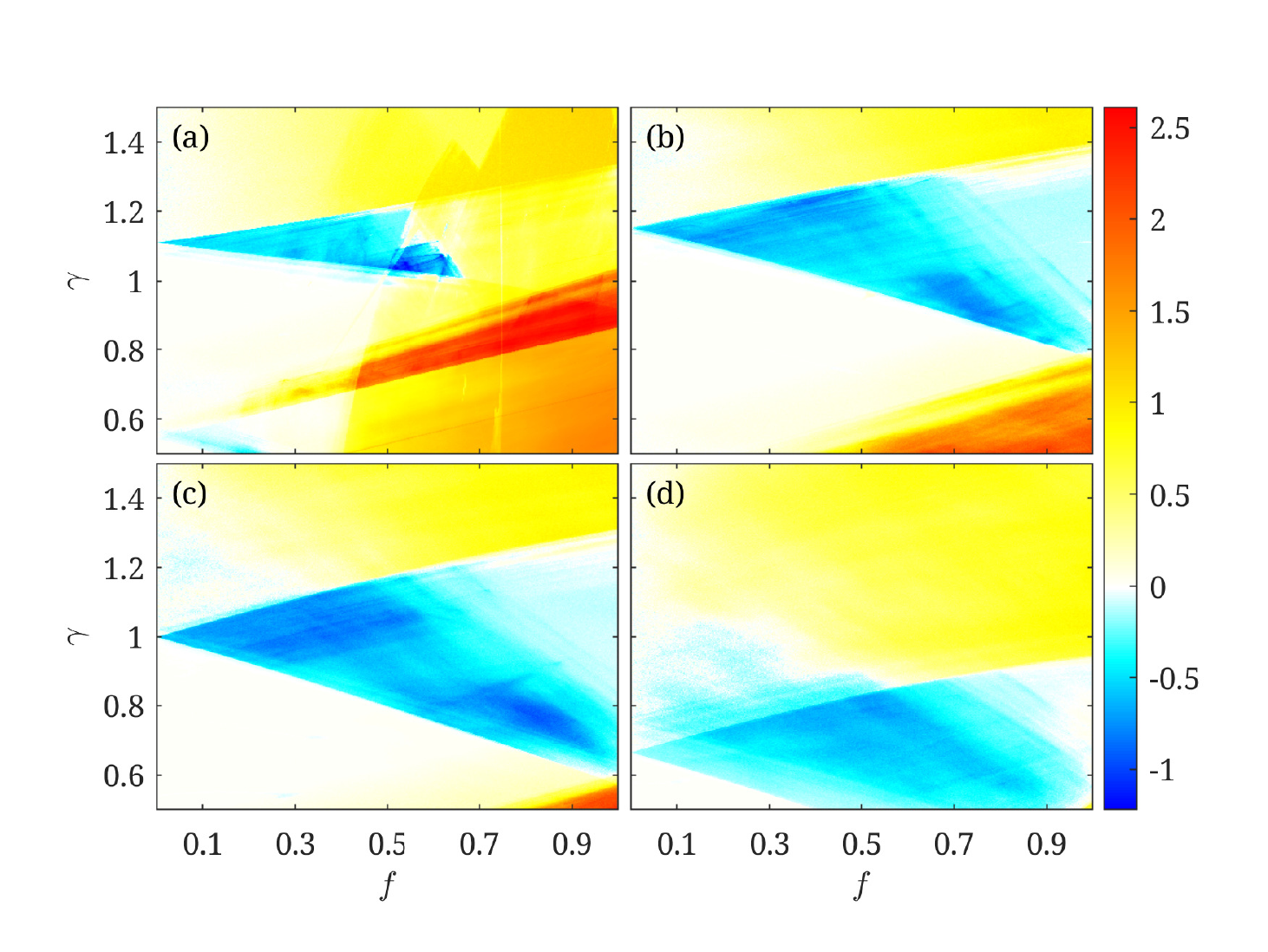}
	\caption{The average velocity $\langle v_\vgamma \rangle$ of the Brownian particle versus the static bias $f$ and friction coefficient $\gamma$ for various $\omega_\vgamma$ with $a=24.4375$ and $D=0$. Panel (a)~$\omega_\vgamma=8.6$, (b)~$\omega_\vgamma=9.3$, (c)~$\omega_\vgamma=10.0$, (d)~$\omega_\vgamma=10.6$.}
	\label{fig:gfw:w}
\end{figure}

In Fig. \ref{fig:gfw:w} we depict the directed transport $\langle v_\vgamma \rangle$ for the same subspace, namely $\langle v_\vgamma \rangle(f,\gamma)$ but for the fixed amplitude $a = 24.4375$ of the harmonic driving and different angular frequencies $\omega_1$ of the latter perturbation. The reader can observe there the evolution of one wedge-like area of absolute negative mobility corresponding to alteration of $\omega_\vgamma$. When $\omega_\vgamma$ increases, this structure is enlarged, with a maximum at $\omega_\vgamma=10.0$, and then starts to disappear. Likewise, there is an optimal value of the static bias for which the absolute negative mobility emerges for the broadest interval of $\gamma$. \emph{If $\omega_\vgamma$ grows the region of negative velocity is moved towards smaller dissipation $\gamma$.} A closer look at Fig. \ref{fig:gfw:w} reveals barely visible areas of negative directed transport that look like copies of the main one, but shifted up or down and dimmed.

The above presented approach to analyze the dynamics of a driven Brownian particle in the multidimensional parameter space by dividing it onto several subspaces allowed us to draw a number of general \emph{qualitative} conclusions regarding the emergence of the absolute negative mobility effect. Firstly, increasing the amplitude of the external perturbation applied to the particle, either static bias $f$ or harmonic driving $a$, causes a shift of the absolute negative mobility regions towards lower values of dissipation $\gamma$. This fact can be seen explicitly in \mbox{Fig. \ref{fig:gaw:a}} or Fig. \ref{fig:gfw:a}. Secondly, when the frequency $\omega_\vgamma$ of the harmonic driving grows the regions of absolute negative mobility displace in the direction of smaller dissipation $\gamma$ and greater constant bias $f$, c.f. Fig. \ref{fig:gfw:w}. Thirdly, in most cases temperature $D$ influences destructively the anomalous transport behavior which we exemplified in Fig. \ref{fig:gaw:D}.
\begin{figure}[t]
	\centering
	\includegraphics[width=1.0\linewidth]{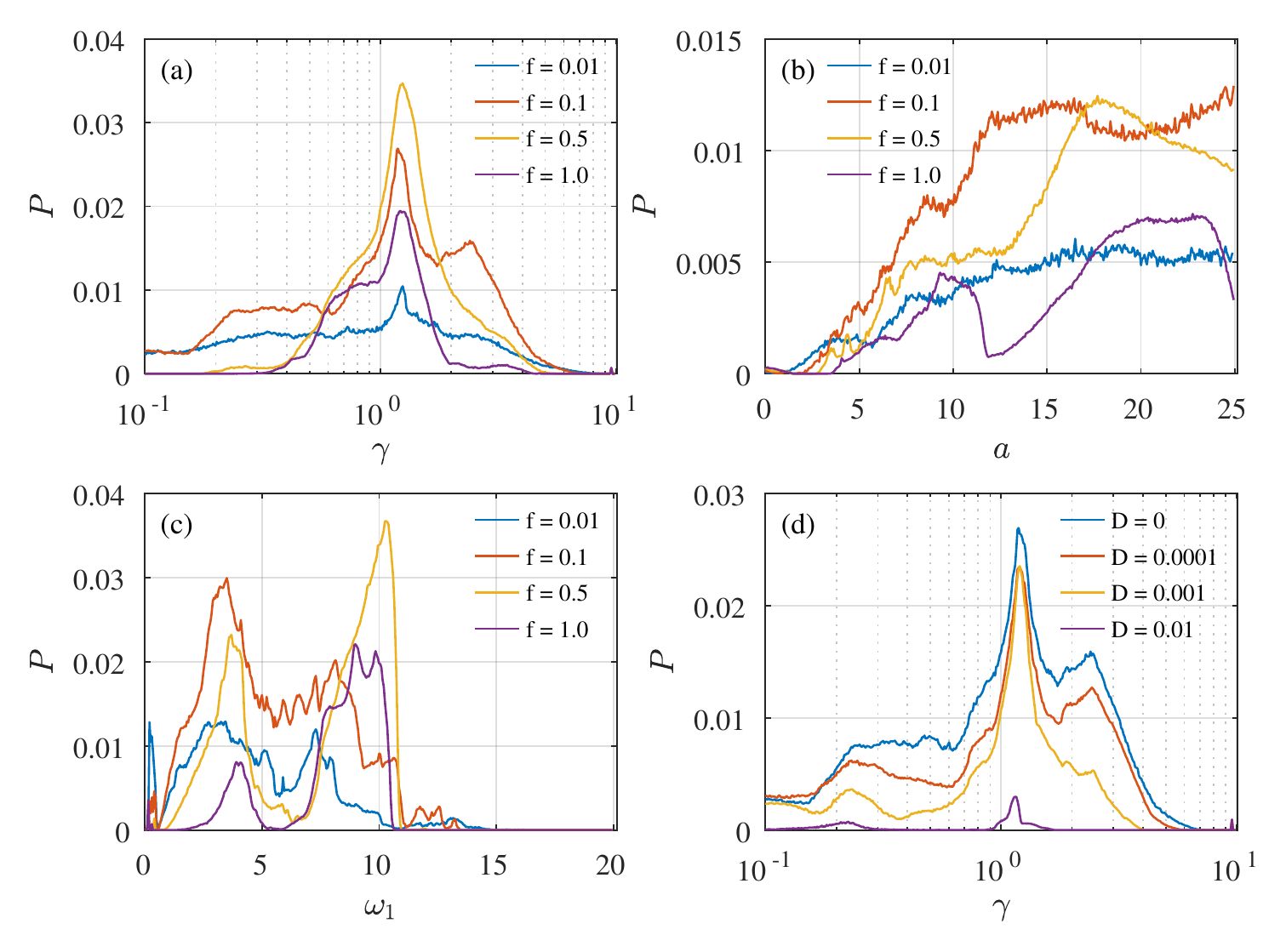}
	\caption{Probability $P$ for the emergence of absolute negative mobility as a function of (a) friction coefficient $\gamma$, (b) amplitude $a$ and (c) frequency $\omega_\vgamma$, for different constant force $f$. On (d) $P$ is plotted versus $\gamma$ for various thermal noise intensities $D$ and $f = 0.1$.} %The threshold for the velocity to be considered negative was -0.1.}
	\label{fig:gaw:eta}
\end{figure}

Exploiting the GPU supercomputers to investigate the absolute negative mobility of the driven Brownian particle let us to attain not only qualitative remarks about the emergence of this effect in the parameter space but also obtain the important \emph{quantitative} results. We depict them in Fig. \ref{fig:gaw:eta} where the fraction of the investigated space for which the absolute negative mobility phenomenon occurs (probability $P$) versus different parameters of the studied model for several values of the static bias $f$ is presented. Panel (a) illustrate the latter quantity as a function of dissipation $\gamma$. One can note that in the both extreme limits $\gamma \to 0$ and $\gamma \to \infty$ the anomalous transport behaviour completely disappears. It is consistent with the state of the art of this paradoxical effect. However, regardless of the magnitude of constant force $f$ there is a common optimal value of $\gamma \approx 1.1749$ for which $P$ is maximal, i.e. the absolute negative mobility emerges most frequently in the parameter space. The reader can observe that the extremum of $P$ is maximal for the bias $f = 0.5$. In panel (b) of the same Fig. we show $P$ versus the amplitude $a$ of the external harmonic driving. For $a=0$ there is no transport in the negative direction which agrees with the statement that the harmonic driving is necessary for this anomalous behaviour to emerge. For all values of $f$ the probability $P$ initially rises when $a$ increases from zero. Then it strongly depends on the value of $f$ and can have both local maxima and minima. However, the general tendency is that for not too large $a$ the absolute negative mobility occurs more frequently when the amplitude $a$ is increased. In panel (c) the dependence of $P$ on the frequency $\omega_\vgamma$ is presented. For all values of the constant force $f$ there are two intervals where $P$ has pronounced extrema. Their locations vary for different $f$, but the first one is in the vicinity of $\omega_\vgamma \approx 3.5$ and the other one is between $\omega_\vgamma \in [7,11]$. In the former region the highest peak is reached for $f=0.1$ while for the latter the optimal value of force is $f=0.5$ for which the global maximum of $P$ is attained when $\omega_\vgamma \approx 10.25$. The higher the static bias $f$, the lower is the cutoff frequency $\omega_\vgamma$ for which the absolute negative mobility ceases to exist. Regardless of the magnitude of the load $f$ the probability $P$ vanishes if $\omega_\vgamma > 15$. Finally, in panel (d) the impact of thermal fluctuations on the distribution $P$ is illustrated for the fixed $f = 0.1$. For most of $\gamma$ thermal noise have destructive influence on the occurrence of absolute negative mobility with the highest value of $P$ in the deterministic case $D = 0$. However, careful inspection of the figure reveals that for small $\gamma$ the probability $P$ for $D=0.0001$ is larger than for $D=0$. This fact testifies that thermal fluctuations indeed can induce the absolute negative mobility of the driven Brownian particle.
\subsection{Second scaling with friction coefficient $\gamma=1$}
For fixed values of dissipation $\gamma$ and inertia $m$ in the first and second scaling, respectively, one is able to transform the results obtained in each scaling to the other one by using the relations presented in Sec. 2.3. Therefore qualitative conclusions about the emergence of absolute negative mobility in the parameter space which we made in the last section hold true provided that the relation $m = 1/\gamma^2$ is taken into account. It basically means that inertia $m$ is inversely proportional to dissipation $\gamma^2$. The exception from the above rule is when the directed transport $\langle v \rangle$ is studied in the parameter plane involving inertia $m$. Then the relation between the frequencies $\omega_\vgamma = \sqrt{m}\omega_\vm$ tells that it is not possible to fix single $\omega_\vm$ in such a way that it would correspond to $\omega_\vgamma$ for all inertia $m$ simultaneously and consequently there is no one-to-one correspondence between the analyzed parameter subspaces. For this reason it is not pointless to investigate the emergence of the absolute negative mobility effect also in the second scaling.
\begin{figure}[t]
	\centering
	\includegraphics[width=1.0\linewidth]{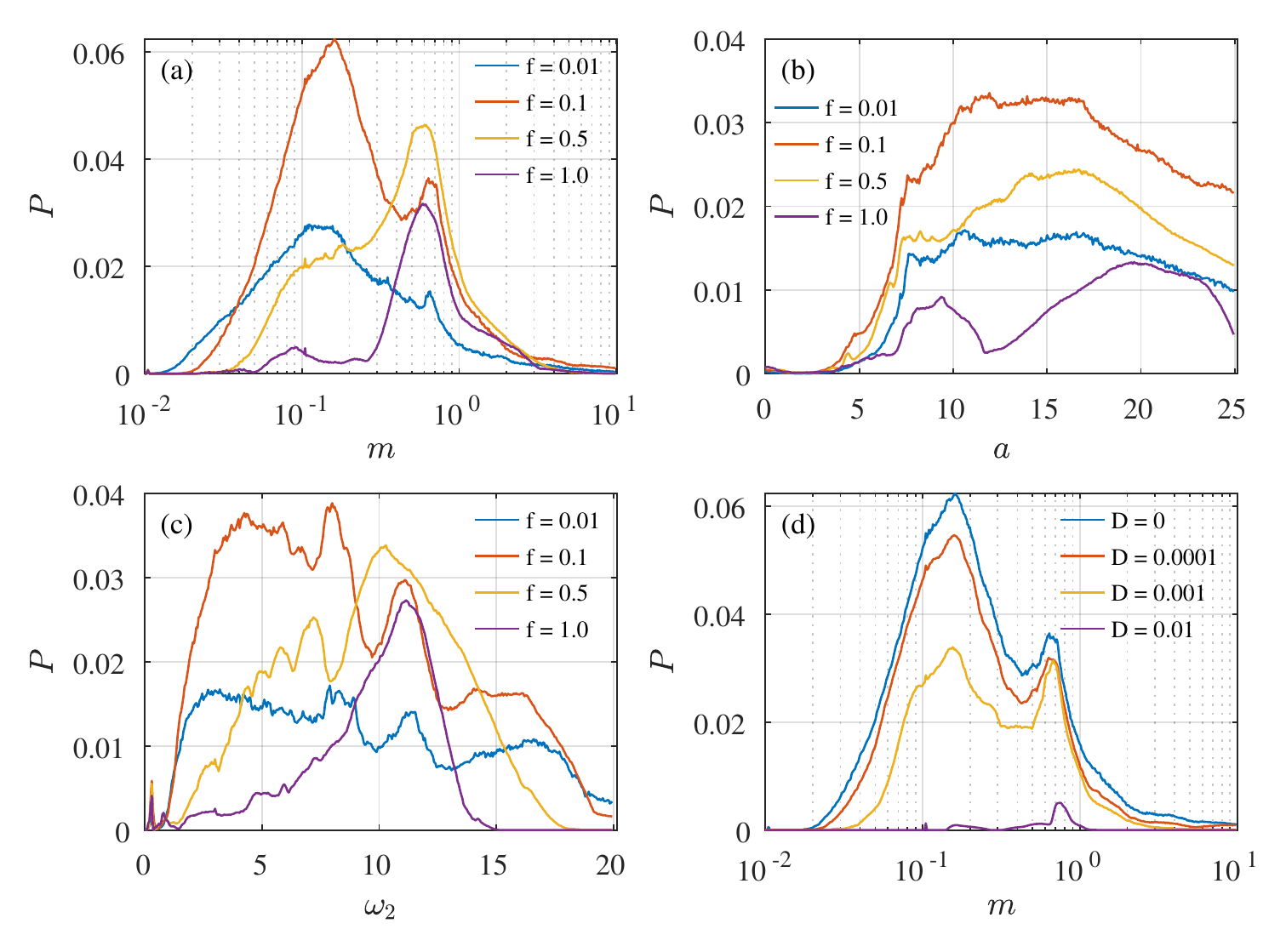}
	\caption{Probability $P$ for the emergence of absolute negative mobility as a function of (a) mass $m$, (b) amplitude $a$ and (c) frequency $\omega_\vm$, for four values of constant force $f$. On (d) $P$ is plotted versus $m$ for different thermal noise intensities $D$ and $f = 0.1$.} %The threshold for the velocity to be considered negative was -0.1.}
	\label{fig:maw:eta}
\end{figure}

In Fig. \ref{fig:maw:eta} we show the probability $P$ of emergence of this anomalous transport behaviour versus different quantity characterizing the studied model. These results are analogous to the corresponding ones depicted in Fig. \ref{fig:gaw:eta} but now they are for the second scaling. In panel (a) $P$ is plotted as a function of the dimensionless mass $m$. The reader can immediately realize that in both extreme limits $m \to 0$ and $m \to \infty$ the absolute negative mobility ceases to exist. Again it is consistent with the current state of 	knowledge. There are two values of $m$ for which the probability $P$ displays pronounced maximum. The first one is $m \approx 0.16$ and the second $m \approx 0.6$. The corresponding optimal forces reads $f = 0.1$ and $f = 0.5$ for the former and latter inertia $m$, respectively. In panel (b) of the same Fig. we investigate the distribution $P$ versus the amplitude $a$ of the harmonic driving. For every static bias $f$ there is a range of $a$ for which the absolute negative mobility emerges most frequently, e.g. for $f = 0.1$ it is $a \in [11, 16]$. It is expected that the probability $P$ will eventually vanish for growing $a$. It is due to the fact that in such a case other perturbations in the dynamics which are necessary for the anomalous transport to arise will be negligible.  The most significant difference between plots of $P$ can be seen on Fig. \ref{fig:maw:eta} (c) and \ref{fig:gaw:eta} (c), where the dependence of $P$ on the angular frequency $\omega$ is captured. The similarity between them is that for each $f$ there is the cutoff frequency $\omega$ above which the absolute negative mobility does not occur. Naturally it is because the dimensionless frequencies $\omega_\vgamma$ and $\omega_\vm$ are proportional to each other. Moreover, minor changes of $\omega$ in both cases can intensify or weaken the emergence of absolute negative mobility. However, in contrast to the panel Fig. \ref{fig:gaw:eta} (c), there exists a common characteristic frequency $\omega_\vm \approx 11$ for which the probability $P$ displays the pronounced maximum for every studied force $f$. Finally, in panel (d) the influence of thermal noise intensity $D$ on the occurrence of absolute negative mobility $P$ is depicted. For the whole range of the investigated inertia $m$ temperature $D$ acts destructively on this anomalous transport feature.
\section{Conclusions}
In this work we revisited the problem of anomalous transport in driven periodic systems. Specifically, we considered a paradigmatic model of nonequilibrium statistical physics, namely, an inertial Brownian particle moving in a symmetric spatially periodic potential which in addition is exposed to both an external harmonic driving and a static load. We focused on the particular instance of anomalous transport behaviour in the form of the counter-intuitive absolute negative mobility effect, in which the net particle movement is opposite to the direction of the applied biasing force around zero load.

It has been under investigation for many years, however, since the studied system possesses an extremely rich five-dimensional parameter space which for the same time has been too complex to explore systematically, research to date has been targeted mostly at the single case studies. In contrast, in the present work we exploited the state of the art computer simulations to analyze the emergence of absolute negative mobility phenomenon in the parameter space with unprecedented resolution of nearly $10^9$ parameter regimes by harvesting the power of GPU supercomputers. This innovative method allowed us to draw a few general qualitative conclusions about the absolute negative mobility. Moreover, the astonishing number of considered regimes made it possible to obtain also a number of important quantitative results. 

In particular, we determined how the occurrence of absolute negative mobility effect changes under the influence of the system parameters. The amplitude of the external perturbation, either static bias $f$ or harmonic driving $a$, induce a shift of the anomalous transport regions towards lower values of dissipation $\gamma$. When the frequency $\omega$ of the harmonic driving grows the occurrence of absolute negative mobility displace in the direction of smaller damping $\gamma$ and greater constant force $f$. Thermal fluctuations $D$ typically have destructive impact on this anomalous transport behaviour. Surprisingly, it turned out that regardless of the magnitude of constant force $f$ there is a common optimal value of dissipation $\gamma \approx 1.1749$ for which the absolute negative mobility emerges most frequently in the parameter space. In contrast, for inertia $m$ the dichotomous behaviour was detected, i.e. depending on the static bias $f$ there are two global optima for this effect. When the constant force $f \approx 0.1$ is significantly smaller than the potential barrier the optimal mass reads $m \approx 0.16$, whereas for the load $f \approx 1.0$ comparable to the potential barrier it reads $m \approx 0.6$. %This suggests the existence of two separate mechanisms that induce the negative mobility and depend on mass $m$ and external bias $f$. 
The general tendency is that the absolute negative mobility occurs more frequently when the amplitude $a$ of the external driving is increased. The effect of external harmonic driving frequency $\omega$ is most complex, however, for many studied loads $f$ the frequency $\omega \approx 11$ seems to be the most optimal choice with respect to the emergence of absolute negative mobility.

Summarizing, our findings for the paradigmatic model of nonequilibrium statistical physics can be straightforwardly corroborated experimentally with a wealth of physical systems including, among others, the Josephson junctions and cold atoms dwelling in optical lattices which nowadays are intensively studied. We believe that the results presented in this work will serve as a platform for both experimenters and theorists to further investigate anomalous transport processes occurring in driven periodic systems.
\section*{Acknowledgment}
This work was supported by the Grant No. NCN 2017/26/D/ST2/00543 (J.S.)


\section*{References}
\begin{thebibliography}{99}
	\bibitem{strogatz} Strogatz S H 2015 \textit{Nonlinear Dynamics and Chaos: With Applications to Physics, Biology, Chemistry and Engineering} (CRC, Boca Raton)
	\bibitem{ott} Ott E 2002 \textit{Chaos in Dynamical Systems} (Cambridge University Press, Cambridge)
	\bibitem{pisarchik2014} Pisarchik A N and Feudel U 2014 \textit{Phys. Rep.} \textbf{540}, 167
	\bibitem{spiechowicz2021pre} Spiechowicz J and {\L}uczka J 2020 \textit{Phys. Rev. E} \textbf{104} 024132
	\bibitem{gammaitoni1998} Gammaitoni L, H\"anggi P, Jung P and Marchesoni F 1998 \textit{Rev. Mod. Phys.} \textbf{70} 223
	\bibitem{metzler2014} Metzler R, Jeon J H, Cherstvy A G and Barkai E 2014 \textit{Phys. Chem. Chem. Phys.} \textbf{16} 24128
	\bibitem{spiechowicz2016scirep} Spiechowicz J, {\L}uczka J, H\"anggi P 2016 \textit{Sci. Rep.} \textbf{6} 30948
	\bibitem{spiechowicz2017scirep} Spiechowicz J and {\L}uczka J 2017 \textit{Sci. Rep.} \textbf{7} 16451
	\bibitem{bialas2020} Bia{\l}as K, {\L}uczka J, H\"anggi P and Spiechowicz J 2020 \textit{Phys. Rev. E} \textbf{102} 042121
	\bibitem{hanggi2009} H\"anggi P and Marchesoni F 2009 \textit{Rev. Mod. Phys.} \textbf{81} 387
	\bibitem{cubero2016} Cubero D and Renzoni F 2016 \textit{Brownian Ratchets: From Statistical Physics to Bio and Nano-motors} (Cambridge, Cambridge University Press)
	\bibitem{jarzynski2011} Jarzynski C 2011 \textit{Annu. Rev. Condens. Matter Phys.} \textbf{2} 329
	\bibitem{campisi2011} Campisi M, H\"anggi P, Talkner P 2011 \textit{Rev. Mod. Phys.} \textbf{83} 771
	\bibitem{talkner2020} Talkner P and H\"anggi P 2020 \textit{Rev. Mod. Phys.} \textbf{92} 041002
	\bibitem{eichhorn2002a} Eichhorn R, Reimann P and H\"anggi P 2002 \textit{Phys. Rev. Lett.} \textbf{88} 190601
	\bibitem{eichhorn2002b} Eichhorn R, Reimann P and H\"anggi P 2002 \textit{Phys. Rev. E.} \textbf{66} 066132
	\bibitem{ros2005} Ros A, Eichhorn R, Regtmeier J, Duong T T, Reimann P and Anselmetti D 2005 \textit{Nature} \textbf{436} 928
	\bibitem{machura2007} Machura {\L}, Kostur M, Talkner P,  {\L}uczka J and H\"anggi P 2007 \textit{Phys. Rev. Lett.} \textbf{98} 40601
	\bibitem{speer2007} Speer D, Eichhorn R and Reimann P 2007 \textit{Phys. Rev. E} \textbf{76} 051110
	\bibitem{kostur2008} Kostur M, Machura L, Talkner P, H\"anggi P and {\L}uczka J 2008 \textit{Phys. Rev. B} \textbf{77} 104509				
	\bibitem{nagel2008} Nagel J et al. 2008 \textit{Phys. Rev. Lett.} \textbf{100} 217001
	\bibitem{eichhorn2010} Eichhorn R, Regtmeier J, Anselmetti D and Reimann P 2010 \textit{Soft Matter} \textbf{6} 1858
	\bibitem{spiechowicz2013jstatmech} Spiechowicz J, {\L}uczka J and H\"anggi P 2013 \textit{J. Stat. Mech} P02044
	\bibitem{malgaretti2014} Malgaretti P, Pagonabarraga I and Rubi J M 2014 \textit{Phys. Rev. Lett.} \textbf{113} 128301
	\bibitem{spiechowicz2014pre} Spiechowicz J, H\"anggi P and {\L}uczka J 2014 \textit{Phys. Rev. E} \textbf{90} 032104
	\bibitem{sarracino2016} Sarracino A, Cecconi F, Puglisi A and Vulpiani A 2016 \textit{Phys. Rev. Lett.} \textbf{117} 174501
	\bibitem{slapik2018} Slapik A, {\L}uczka J and Spiechowicz J 2018 \textit{Commun. Nonlinear Sci. Numer. Simulat.} \textbf{55} 316
	\bibitem{cecconi2018} Cecconi F, Puglisi A, Sarracino A and Vulpiani A 2018 \textit{J. Phys. Condens. Matter} \textbf{30} 264002
	\bibitem{ai2018} Ai B-Q, Zhu W, He Y and Zhong W 2018 \textit{J. Chem. Phys} \textbf{149} 164903
	\bibitem{mukamel2018} Cividini J, Mukamel D and Posch H A 2018\textit{J. Phys. A: Math. Theor.} \textbf{51} 085001
	\bibitem{slapik2019prl} Slapik A, {\L}uczka J, H\"anggi P and Spiechowicz J 2019 \textit{Phys. Rev. Lett.} \textbf{122} 070602
	\bibitem{slapik2019prappl} Slapik A, {\L}uczka J, Spiechowicz J 2019 \textit{Phys. Rev. Appl.} \textbf{12} 054002
	\bibitem{spiechowicz2019njp} Spiechowicz J, H\"anggi P and {\L}uczka J 2019 \textit{New J. Phys} \textbf{21} 083029
	\bibitem{ai2020physa} Li J-J, Xie H-Z, Li T-Ch, Ai B-Q 2020 \textit{J. Stat. Mech.} \textbf{560} 125164
	\bibitem{ai2020} Luo Y, Zeng Ch and Ai B-Q 2020 \textit{Phys. Rev. E} \textbf{102} 042114
	\bibitem{ai2022} Wu J-Ch, Lin F-J and Ai B-Q 2022 \textit{Soft Matter} \textbf{18} 1194
	\bibitem{spiechowicz2015cpc} Spiechowicz J, Kostur M and Machura {\L} 2015 \textit{Comp. Phys. Commun.} \textbf{191} 140
	\bibitem{marconi2008} Marconi U M B, Puglisi A, Rondoni L and Vulpiani A 2008 \textit{Phys. Rep} \textbf{461} 111
	\bibitem{fulde1975} Fulde P, Pietronero L, Schneider W R and Str\"assler S 1975 \textit{Phys. Rev. Lett.} \textbf{35} 1776
	\bibitem{dieterich1980} Dieterich W, Fulde P and Peschel I 1980 \textit{Adv. Phys.} \textbf{29} 527-605
	\bibitem{langevin} Coffey W T, Kalmykov Y P, and Waldron J T 2004 \textit{The Langevin
Equation} (World Scientific, Singapore, 2004); see Secs. 5 and 7-10 therein
	\bibitem{gruner1981} Gr\"uner G, Zawadowski A and Chaikin P M 1981 \textit{Phys. Rev. Lett.}
\textbf{46} 511
	\bibitem{kautz} Kautz R L 1996 \textit{Rep. Prog. Phys.} \textbf{59} 935
%	\bibitem{spiechowicz2014prb} Spiechowicz J, H\"anggi P and {\L}uczka J 2014 \textit{Phys. Rev. B} \textbf{90} 054520
	\bibitem{blackburn2016} Blackburn JA, Cirillo M, Gronbech-Jensen N 2016 \textit{Phys. Rep.} \textbf{611} 1
	\bibitem{spiechowicz2015chaos} Spiechowicz J and {\L}uczka J 2015 \textit{Chaos} \textbf{25} 053110
%	\bibitem{spiechowicz2015njp} Spiechowicz J and {\L}uczka J 2015 \textit{New J. Phys.} \textbf{17} 023054
	\bibitem{spiechowicz2019chaos} Spiechowicz J and {\L}uczka J 2019 \textit{Chaos} \textbf{29} 013105
	\bibitem{lutz2013} Lutz E and Renzoni F 2013 \textit{Nat. Phys.} \textbf{9} 615
	\bibitem{lutz2017} Kindermann F et al. 2017 \textit{Nat. Phys.} \textbf{13} 137
	\bibitem{hanggi2020} H\"anggi P, {\L}uczka J and Spiechowicz J 2020 \textit{Acta Phys. Polon. B} \textbf{51} 1131
	\bibitem{spiechowicz2021pre2} Spiechowicz J and {\L}uczka J 2021 \textit{Phys. Rev. E} \textbf{104} 034104
	\bibitem{denisov2014} Denisov S, Flach S and H\"anggi P 2014 \textit{Phys. Rep.} \textbf{538} 77
	\bibitem{platen} Platen E and Bruti-Liberati N 2010 \textit{Numerical Solution of Stochastic Differential Equations with Jumps in Finance} (Stochastic Modelling and Applied Probability) (Berlin, Springer-Verlag)
\end{thebibliography}
\end{document}